\documentclass[11pt,a4paper]{article}
\usepackage{jcappub}
\usepackage[utf8]{inputenc}

\title{Perturbations of Cosmological and Black Hole Solutions in Massive
gravity and Bi-gravity}

\author[a]{Tsutomu Kobayashi,}
\author[b]{Masaru Siino,}
\author[b]{Masahide Yamaguchi,}
\author[b]{Daisuke Yoshida}

\affiliation[a]{Department of Physics, Rikkyo University, Toshima, Tokyo
175-8501, Japan}

\affiliation[b]{Department of Physics, Tokyo Institute of Technology,
Tokyo 152-8551, Japan}

\emailAdd{tsutomu@rikkyo.ac.jp}
\emailAdd{msiino@th.phys.titech.ac.jp}
\emailAdd{gucci@phys.titech.ac.jp}
\emailAdd{yoshida@th.phys.titech.ac.jp}

\abstract{We investigate perturbations of a class of spherically
symmetric solutions in massive gravity and bi-gravity. The background equations of
motion for the particular class of solutions we are interested in
reduce to a set of the Einstein equations with a cosmological constant.
Thus, the solutions in this class include all
the spherically symmetric solutions in general relativity,
such as the Friedmann-Lema\^{i}tre-Robertson-Walker solution and the
Schwarzschild (-de Sitter) solution,
though
the one-parameter family of two parameters of the theory
admits such a class of solutions.
We find that the equations of
motion for the perturbations of this class of solutions also reduce to the
perturbed Einstein equations at first and second order. Therefore, the stability
of the solutions coincides with that of the corresponding solutions in general relativity.
In particular, these solutions do not suffer from non-linear instabilities
which often appear in the other cosmological solutions in massive
gravity and bi-gravity.}

\begin{document}
\maketitle

\section{Introduction}
\label{Sec.1}

Massive gravity is one of the potent candidates for modified theory of
general relativity.  As early as in 1939, a linear theory of massive
gravity was proposed by Fierz and Pauli (FP)~\cite{Fierz:1939ix}. In
order to avoid the inconsistency on massless limit of this
theory~\cite{vanDam:1970vg,Zakharov:1970cc}, a nonlinear extension of
the FP theory was considered~\cite{Vainshtein:1972sx}.  Boulware and
Deser, however, found that the nonlinear theory simply extended from the
FP theory contains an unphysical ghost degree of freedom (BD
ghost)~\cite{Boulware:1973my}.  Because of this ghost problem, a healthy
theory of non-linear massive gravity had not been established for a long
time.

Recently, de Rham, Gabadadze, and Tolley (dRGT) proposed a mass
potential which can remove the BD ghost mode in a decoupling
limit~\cite{deRham:2010ik,deRham:2010kj}, and Hassan and Rosen have
finally proven that the dRGT massive gravity theory is free from the BD
ghost without taking the decoupling
limit~\cite{Hassan:2011hr,Hassan:2011tf,Hassan:2011ea}. The dRGT theory of massive
gravity has three parameters, graviton mass $m$ and coupling constants
of nonlinear self interactions, $\alpha_3,\alpha_4$.  Complementary
approaches of the BD ghost problem are studied
in refs.~\cite{deRham:2011rn,Mirbabayi:2011aa,Kugo:2014hja,Gao:2014ula}.

Massive gravity includes a non-dynamical tensor field called fiducial
metric, $f_{\mu\nu}$, in order to construct a mass potential. For example,
the FP and the original dRGT theories are constructed by adopting the
Minkowski fiducial metric. Hassan and Rosen proposed an extended theory
of dRGT massive gravity with a fiducial metric being dynamical as well
by introducing the Einstein-Hilbert term of the fiducial metric in the
action. They proved that this theory is also free from the BD
ghost~\cite{Hassan:2011zd}. Since this theory contains two symmetric dynamical
tensor fields of metrics, it is called bi-gravity theory.

The tests of massive gravity and bi-gravity using cosmological and black
hole solutions have been explored intensively.
In dRGT massive gravity, several types of exact, homogeneous, and
isotropic solutions have been known so far. One example is the open
Friedmann-Lema\^{i}tre-Robertson-Walker (FLRW) solution with the flat
fiducial metric in the FLRW slice~\cite{Gumrukcuoglu:2011ew}. The
second-order perturbations of this solution show nonlinear instability and
hence this solution is not viable
unfortunately~\cite{Gumrukcuoglu:2011zh,DeFelice:2012mx,Pereira:2015jua}. It should be
noted that similar solutions with an anisotropic fiducial metric are
known to be stable~\cite{Gumrukcuoglu:2012aa,DeFelice:2013awa,DeFelice:2013bxa}.
Another example of cosmological solutions is that with a fiducial metric
which is flat but expressed in terms of nontrivial coordinates. This class of
solutions can be divided into two types. The first type includes the
solutions found in
refs.~\cite{Koyama:2011xz,D'Amico:2011jj,Gratia:2012wt}, which exist for
the whole parameter region of $\alpha_3$ and $\alpha_4$, while the other
type includes the solutions found in
refs.~\cite{Chamseddine:2011bu,Kobayashi:2012fz}, which exists only for a
one parameter family in the parameter space $(\alpha_3,\alpha_4)$. Though
the perturbations of the former type of solutions have already been
studied in
refs.~\cite{D'Amico:2012pi,Wyman:2012iw,Khosravi:2013axa,Motloch:2014nwa},
the perturbations of the latter type of solutions have not yet been
investigated. Therefore, in this paper, we focus on the latter type of solutions.
It should be noted that cosmological solutions with non-flat fiducial
metrics are also studied in
refs.~\cite{Langlois:2012hk,Langlois:2013cya,Fasiello:2012rw,Pan:2015eta}.

The situations on cosmological solutions in bi-gravity are
similar. The cosmological solutions found thus far are divided into two
classes~\cite{Volkov:2011an,Volkov:2012cf,Volkov:2012zb,Volkov:2013roa}.
The first class is a solution with diagonal
metric tensors~\cite{vonStrauss:2011mq,Comelli:2011zm,Konnig:2013gxa}. The
perturbations of this class of solutions have been studied
intensively~\cite{Comelli:2012db,Berg:2012kn,Khosravi:2012rk,Comelli:2014bqa,Konnig:2014dna,DeFelice:2014nja,Solomon:2014dua,Konnig:2014xva,Lagos:2014lca,Cusin:2014psa,Enander:2015vja,Akrami:2015qga,Fasiello:2015csa}. On
the other hand, for the other class of a solution with off-diagonal
components of physical or fiducial metric tensor, the perturbations have not yet
been studied, similarly to the case of massive gravity. It should be
noticed that coupling between matter and (bi-)metrics is a nontrivial
issue in bi-gravity and is studied in
refs.~\cite{deRham:2014naa,deRham:2014fha,Gumrukcuoglu:2014xba,Solomon:2014iwa,Gao:2014xaa,Gumrukcuoglu:2015nua,Comelli:2015pua,Heisenberg:2015wja}.

A lot of static and spherically symmetric solutions have also been found up to
now. The classification of such spherically symmetric solutions is
studied in refs.~\cite{Volkov:2012wp,Volkov:2013roa,Volkov:2014ooa}. The
exact Schwarzschild (-de Sitter) solutions are classified to the
following three classes.  The first class is a solution with diagonal metric tensors 
\cite{Comelli:2011wq,Katsuragawa:2013lfa,Ghosh:2015cva,Babichev:2015xha} and linear perturbations of
this class of solutions are studied in
ref.~\cite{Babichev:2013una,Brito:2013wya,Brito:2013yxa,Katsuragawa:2014hda,Babichev:2015xha}
in the framework of both massive gravity and bi-gravity. The second class of
solutions is a solution with a off-diagonal metric tensor and arbitrary $\alpha_3$ and
$\alpha_4$~\cite{Koyama:2011yg}, and the perturbation of this solution
is studied in ref.~\cite{Babichev:2014oua,Babichev:2015xha}.
The last class is a solution with a off-diagonal metric tensor and a special choice of the parameters $\alpha_3$
and
$\alpha_4$~\cite{Nieuwenhuizen:2011sq,Berezhiani:2011mt,Arraut:2013bqa},
where linear perturbations have been studied only in massive gravity
with a flat fiducial metric~\cite{Kodama:2013rea} and not in bi-gravity.

In the present article, we will give
a unified and general treatment for solutions with
an off-diagonal metric tensor in massive gravity and bi-gravity
belonging to a one parameter family of $\alpha_3$ and
$\alpha_4$, which include both
cosmological~\cite{Chamseddine:2011bu,Kobayashi:2012fz} and spherically
symmetric black hole
solutions~\cite{Nieuwenhuizen:2011sq,Berezhiani:2011mt,Volkov:2012wp,Arraut:2013bqa,Kodama:2013rea}. We will
find that the equations of motion for this class of solutions exactly
reduce to those of general relativity (GR) with a cosmological constant
not only at the background and linear (first-order) perturbation level
but also at the level of quadratic (second-order) perturbations. This
result shows that massive gravity and bi-gravity can allow any
spherically symmetric solution of GR including its stability, the evolution
of linear perturbations, and the backreaction from linear perturbations,
while it simultaneously implies that one cannot distinguish massive
gravity or bi-gravity from GR by using spherically symmetric solutions
and their perturbations at least up to quadratic order.

Our paper is organized as follows. In the next section, we briefly
review the theory of bi-gravity (and massive gravity as a trivial case
of a fixed fiducial metric) and derive the equations of motion in a
general setting. In section 3, we derive a generic non-diagonal
spherically symmetric background solution. Then, we investigate linear
perturbations around those background solutions in section 4. There
we will see that the terms coming from the mass potential must vanish in
order to satisfy the Bianchi identity. In section 5, we investigate
higher order perturbations and find that the same results as the linear
perturbations apply for the quadratic perturbations. The final section is
devoted to summary and discussion. Some details will be given in the
appendices.

\section{Review of bi-gravity}

In this section, we give a brief review of bi-gravity.
Bi-gravity is a theory consisting of two dynamical tensor fields,
$g_{\mu\nu}$ and $f_{\mu\nu}$, called physical and fiducial metrics, respectively.
Massive gravity can be understood as a special case where the fiducial metric is fixed and non-dynamical.
Its action is given by the Einstein-Hilbert term for each metric with the
interaction term $S_{\rm mass}$ and matter actions:
\begin{eqnarray}
&& S = \frac12 M_{\rm pl}^2\int d^4 x \sqrt{-g} R[g]
 + \frac12 \kappa^2 M_{\rm pl}^2 \int d^4x \sqrt{-f} R[f] \nonumber \\
&& \qquad \qquad +S_{\rm mass}[g,f] 
   + S_{\rm matter}[g] + S_{\rm matter}[f],
\end{eqnarray}
where $R[\cdot]$ is the Ricci scalar and $\kappa$ represents the ratio
of the effective Planck masses for $g_{\mu\nu}$ and $f_{\mu\nu}$. In the case of
$\kappa = 0$, the tensor field $f_{\mu\nu}$ does not have its kinetic term
and hence is non-dynamical. This case corresponds to the massive gravity theory
originally proposed by de Rham, Gabadadze, and Tolley~\cite{deRham:2010ik,deRham:2010kj}.
$S_{\rm matter}[g]$ and $S_{\rm matter}[f]$ are the matter actions coupled to
$g$ and $f$, respectively,
\begin{eqnarray}
 S_{\rm matter}[g] &=& \int d^4 x \sqrt{-g}~{\cal L}_{\rm matter}^{(g)}, \\
 S_{\rm matter}[f] &=& \int d^4 x \sqrt{-f}~{\cal L}_{\rm matter}^{(f)}.
\end{eqnarray}
Here we implicitly assume the matter actions possess the two general covariance with respect to $g_{\mu\nu}$ and $f_{\mu\nu}$ separately, though the full theory does not have such a symmetry.
It should be noted that another type of matter coupling, which does not possess the two general covariance, is also studied by \cite{deRham:2014naa,deRham:2014fha,Gumrukcuoglu:2014xba,Solomon:2014iwa,Gao:2014xaa,Gumrukcuoglu:2015nua,Comelli:2015pua,Heisenberg:2015wja}.
The energy-momentum tensors coming from $S_{\rm matter}[g]$ and $S_{\rm matter}[f]$ are defined as
\begin{eqnarray}
 T_{(g)}{}^{\mu}{}_{\nu} &=& \frac{2}{\sqrt{-g}} g^{\mu\rho}\frac{\delta S_{\rm
  matter}[g]}{\delta g^{\rho\nu}}, \\
 T_{(f)}{}^{\mu}{}_{\nu} &=& - \frac{2}{\sqrt{-f}} \frac{\delta S_{\rm
  matter}[f]}{\delta f_{\mu\rho}}f_{\rho\nu}.
\end{eqnarray}
Due to the two general covariance of matter actions we assumed,
both energy-momentum tensors are conserved, that is, $\nabla^{(g)}_{\mu} T_{(g)}{}^{\mu}{}_{\nu} = 0$ and
$\nabla^{(f)}_{\mu} T_{(f)}{}^{\mu}{}_{\nu} = 0$, where $\nabla^{(g)}_{\mu}$ and
$\nabla^{(f)}_{\mu}$ are the covariant derivatives with respect to
$g_{\mu\nu}$ and $f_{\mu\nu}$, respectively. Hereafter, we will omit the
suffixes $g$ and $f$ when no confusion is expected.

Now the interaction term $S_{\rm mass}$ is tuned to be free from the BD ghost
mode and given by
\begin{eqnarray}
S_{\rm mass}[g,f]
&=&\frac{M_{\rm pl}^2}{2}\int d^4x \sqrt{-g}\, 2m^2
\sum_{i=0}^{4}\beta_i e_i(\gamma),
\end{eqnarray}
where $i$-th order contributions $e_i(\gamma)$ are
given by
\begin{eqnarray}
e_0(\gamma)&=&1,\\
e_1(\gamma)&=&{\rm Tr}[\gamma],\\
e_2(\gamma)&=&\frac{1}{2}\left(
{\rm Tr}[\gamma]^2-{\rm Tr}[\gamma^2]
\right),\\
e_3(\gamma)&=&
\frac{1}{3!}\left(
{\rm Tr}[\gamma]^3-3{\rm Tr}[\gamma]{\rm Tr}[\gamma^2]+2{\rm Tr}[\gamma^3]
\right),\\
e_4(\gamma)&=& \det(\gamma),
\end{eqnarray}
with
\begin{eqnarray}
\gamma^{\mu}{}_{\nu} = (\sqrt{g^{-1}f})^{\mu}{}_{\nu},
\end{eqnarray}
and $m, \beta_i$ being free parameters of the interaction term.
Since they
always appear in the combination $m^2 \beta_i$, essentially there are
five free parameters. The space of parameters corresponds to the one of the three
parameters of the dRGT theory, $m, \alpha_3, \alpha_4$, and two
cosmological constants, $\Lambda^{(g)}, \Lambda^{(f)}$, and their relations are
given by\footnote{ It is useful to rewrite the action in term of these
parameters as follows,
\begin{eqnarray*}
 S_{\rm mass}[g,f]
&=&\frac{M_{\rm pl}^2}{2}\int d^4 x \left[\sqrt{-g}2m^2
\sum_{i=2}^{4}\alpha_i e_i({\cal K})
+ \sqrt{-g}(-2\Lambda^{(g)})
+ \sqrt{-f}(-2\kappa^2\Lambda^{(f)})\right],
\end{eqnarray*}
with $ {\cal K}^{\mu}{}_{\nu} = \delta^{\mu}{}_{\nu} -\gamma^{\mu}{}_{\nu} $ and $\alpha_2 = 1$.}
\begin{eqnarray}
m^2\beta_0&=&-\Lambda^{(g)}+m^2(6+4\alpha_3+\alpha_4),\\
m^2\beta_1&=&m^2(-3-3\alpha_3-\alpha_4),\\
m^2\beta_2&=&m^2(1+2\alpha_3+\alpha_4),\\
m^2\beta_3&=&m^2(-\alpha_3-\alpha_4),\\
m^2\beta_4&=&-\kappa^2 \Lambda^{(f)} + m^2 \alpha_4.
\end{eqnarray}

Taking the variation of the action with respect to $g_{\mu\nu}$ and $f_{\mu\nu}$, we will obtain the
equations of motion for the two tensor fields. The
equations of motion for $g_{\mu\nu}$ are given by
\begin{eqnarray}
G[g]^{\mu}{}_{\nu}+X_{(g)}{}^{\mu}{}_{\nu}= \frac{1}{M_{\rm pl}^2} T_{(g)}{}^{\mu}{}_{\nu},
\end{eqnarray}
where $G[g]^{\mu}{}_{\nu}$ is the Einstein tensor constructed from $g_{\mu\nu}$ and
\begin{eqnarray}
X_{(g)}{}^{\mu}{}_{\nu} &=& 2m^2
\left(
\tau^{\mu}{}_{\nu}
-
\frac{1}{2}\delta^{\mu}_{\nu}\sum_{i=0}^{3}\beta_i e_i(\gamma)
\right),\\
\tau^{\mu}{}_{\nu}
&=&
\frac{1}{2}\left[
\beta_1\gamma^{\mu}_{\ \nu} 
+\beta_2\left(e_1(\gamma)\gamma^{\mu}_{\ \nu}-(\gamma^2)^{\mu}_{\ \nu}\right)\right.\notag\\&&\left.
+\beta_3\left(e_2(\gamma)\gamma^{\mu}_{\ \nu}-e_1(\gamma)(\gamma^2)^{\mu}_{\ \nu}+(\gamma^3)^{\mu}_{\ \nu}\right)
\right].
\end{eqnarray}
The indices here are raised or lowered by $g_{\mu\nu}$.
The equations of motion for $f_{\mu\nu}$ are given by
\begin{eqnarray}
G[f]^{\mu}_{\ \ \nu}+X_{(f)}{}^{\mu}{}_{\nu}
&=& \frac{1}{\kappa^2M_{\rm pl}^2} T_{(f)}{}^{\mu}{}_{\nu},
\end{eqnarray}
where $G[f]{}^{\mu}{}_{\nu}$ is the Einstein tensor constructed from $f_{\mu\nu}$ and
\begin{eqnarray}
X_{(f)}{}^{\mu}_{\ \ \nu}&=&-\frac{m^2}{\kappa^2}\mathrm{sgn}(\det \gamma)\left(\frac{2}{\det{\gamma}}
\tau^{\mu}_{\ \nu}+\beta_4\delta^\mu{}_{\nu}\right).
\end{eqnarray}
The indices here are raised or lowered by $f_{\mu\nu}$.

\section{Bi-spherically symmetric background solutions}
\label{Sec.3}

Here, we attempt to classify some of the spherically symmetric solutions
in bi-gravity and identify those which obey the same equations of
motion as in general relativity.  These classes of solutions include the
cosmological and black hole solutions known so
far~\cite{Chamseddine:2011bu,Volkov:2011an,Volkov:2012cf,Volkov:2012zb,Volkov:2013roa,Volkov:2012wp,Kodama:2013rea}.

Let us consider the following
 bi-spherically symmetric metrics:
\begin{eqnarray}
\bar{g}_{\mu\nu}dx^\mu dx^\nu
&=&\bar{g}_{00}(t,r)dt^2 + 2\bar{g}_{01}(t,r)dtdr + \bar{g}_{11}(t,r)dr^2+R(t,r)^2 d\Omega^2 ,\label{gbar}\\
\bar{f}_{\mu\nu}dx^\mu dx^\nu
&=&\bar{f}_{00}(t,r)dt^2+2\bar{f}_{01}(t,r)dt dr + \bar{f}_{11}(t,r)dr^2 +A^2(t,r)R^2(t,r)d\Omega^2,\label{fbar}
\end{eqnarray}
with $d \Omega^2 = d\theta^2+\sin^2\theta d\phi^2$.
The matrix $\bar{g}^{-1}\bar{f}$ takes the following form,
\begin{eqnarray}
(\bar{g}^{-1}\bar{f})^\mu{}_{\nu}
=
\begin{pmatrix}
 (\bar{g}^{-1}\bar{f})^0{}_{0}&(\bar{g}^{-1}\bar{f})^0{}_{1}&0&0\\
 (\bar{g}^{-1}\bar{f})^1{}_{0}&(\bar{g}^{-1}\bar{f})^1{}_{1}&0&0\\
 0&0&A^2(t,r)&0\\
 0&0&0&A^2(t,r) 
\end{pmatrix},
\end{eqnarray}
and from these ansatz it is straightforward to see
that the square root of the above matrix
is of the form
\begin{eqnarray}
\bar{\gamma}^\mu{}_{\nu}&=&\left(\sqrt{\bar{g}^{-1}\bar{f}}\,\right)^{\mu}{}_{\nu}=
\begin{pmatrix}
 a(t,r)&b(t,r)&0&0\\
 c(t,r)&d(t,r)&0&0\\
 0&0&A(t,r)&0\\
 0&0&0&A(t,r) 
\end{pmatrix} \label{gammabar}.
\end{eqnarray}
It should be emphasized that the following discussion does not rely on
the concrete expressions of $a(t,r)$, $b(t,r)$, $c(t,r)$, and $d(t,r)$,
but rather relies only on the fact that $\bar\gamma^\mu{}_\nu$ is of the form of eq.~(\ref{gammabar}).

As explained earlier, we are interested in the case where the equations
of motion for both metrics reduce to the Einstein equations with
cosmological constants at the background level. Therefore, in order for
$X_{(g)}{}^{\mu}{}_{\nu}$ to be a cosmological term, the non-trivial
off-diagonal components,
\begin{eqnarray}
 \bar{X}_{(g)}{}^{0}{}_{1} &=& -m^2 b [3-2A + (A-3)(A-1)  \alpha_3+(A-1)^2  \alpha_4],\label{Xtr}\\
  \bar{X}_{(g)}{}^{1}{}_{0} &=&-m^2 c[3-2A + (A-3)(A-1)  \alpha_3+(A-1)^2  \alpha_4],\label{Xrt}
\end{eqnarray}
must vanish. We focus on the case of $b(t,r)\ne0$ or $c(t,r)\neq 0$, and $A\neq 1$,
since with
$b(t,r)=c(t,r)=0$, we will obtain a diagonal metrics as mentioned in
Sec.~\ref{Sec.1} and the perturbations of such diagonal solutions have already been studied.

For non-diagonal solutions we are interested in, the condition that
eqs.~(\ref{Xtr}) and~(\ref{Xrt}) vanish leads to
\begin{eqnarray}
A(t,r)=\frac{2 \alpha_3+\alpha_4+1\pm \sqrt{\alpha_3^2+\alpha_3-\alpha_4+1}}{\alpha_3+\alpha_4}={\rm const}.
\label{alpha4equal}
\end{eqnarray}

Another requirement necessary for $\bar{X}_{(g)}{}^{\mu}{}_{\nu}$ to be a
cosmological term is
\begin{eqnarray}
 \bar{X}_{(g)}{}^{0}{}_{0}-\bar{X}_{(g)}{}^{2}{}_{2} =
  (1-A)C(t,r) \left[A-2+(A-1)\alpha_3\right] = 0,\label{X0tt-X033}
\end{eqnarray}
where, we have defined $C(t,r)$ as  
\begin{eqnarray}
C(t,r) = 
m^2
\frac{A^2-A(a+d)+ad-bc}{(1-A)^2}.\label{Cr}
\end{eqnarray}

With $C(t,r) = 0$, at least three eigenvalues of
$\bar{\gamma}^{\mu}{}_{\nu}$ are equal to $A$. This class of solutions
includes the cosmological solutions found in
refs.~\cite{Koyama:2011xz,D'Amico:2011jj,Gratia:2012wt}
and the Schwarzschild solutions obtained in ref.~\cite{Babichev:2014oua}.
The perturbations of those solutions have already been studied in
detail in refs.~\cite{D'Amico:2012pi,Wyman:2012iw,Khosravi:2013axa,Motloch:2014nwa}.

In the present study, we therefore concentrate on the case with $C(t,r) \neq 0$.  In this case, the
solution to eq.~(\ref{X0tt-X033}) is
\begin{eqnarray}
  A=\frac{2+\alpha_3}{1+\alpha_3}.
  \label{alpha3equal}
\end{eqnarray}
Here we have assumed that $\alpha_3\neq -1$.
Equations~(\ref{alpha4equal}) and~(\ref{alpha3equal}) are consistent provided that
the parameters of the theory, $\alpha_3$ and $\alpha_4$, satisfy
\begin{eqnarray}
0=1+\alpha_3+\alpha_3^2-\alpha_4\;\left(=\beta_2^2-\beta_1\beta_3\right).\label{alpha3alpha4}
\end{eqnarray}
This is equivalent to the condition that the two branches of
the solution~(\ref{alpha4equal}) degenerate.
Thus, we see that only the particular one-parameter
family of $\alpha_3$ and $\alpha_4$
 satisfying~(\ref{alpha3alpha4})
admits the class of solutions we are focusing on.
Note that when eq.~(\ref{alpha3alpha4}) is fulfilled $A$ can also be expressed simply as $A=-\beta_2/\beta_3$.
Note also that the
range of $\alpha_4$ is limited as $\alpha_4=(\alpha_3+1/2)^2+3/4 \geq 3/4$.

In this one-parameter
family of $\alpha_3$ and $\alpha_4$ with eq.~(\ref{alpha3alpha4}), the
interaction terms for bi-spherically symmetric metrics (\ref{gbar}) and
(\ref{fbar}) with eq.~(\ref{alpha3equal})
  are of the form of a cosmological term. For $g_{\mu\nu}$
the interaction term gives
\begin{eqnarray}
\bar{X}_{(g)}{}^{\mu}{}_{\nu} &=& 
\Lambda_{\rm eff}^{(g)}\delta^{\mu}{}_{\nu},\label{eomgbar}
\end{eqnarray}
with
\begin{eqnarray}
 \Lambda_{\rm eff}^{(g)} &=& m^2(A-1) +\Lambda^{(g)},\label{Lambdagbar}
\end{eqnarray}
while for $f_{\mu\nu}$
\begin{eqnarray}
 \bar{X}_{(f)}{}{}^{\mu}{}_{\nu}
&=&
\Lambda^{(f)}_{\rm eff}\delta^{\mu}{}_{\nu},\label{Xbarf}\\
\Lambda_{\rm eff}^{(f)}
&=&\mathrm{sgn}(ad-bc)\left(
		       -\frac{m^2}{\kappa^2}\frac{A-1}{A}+\Lambda^{(f)}
		       \right).\label{Xefff}
\end{eqnarray}
In the case of dRGT massive gravity, $f_{\mu\nu}$ is not a
dynamical but a fixed metric, and hence we need not consider the
equations of motion for $f_{\mu\nu}$.

The class of solutions with $C(t,r) \neq 0$ includes the cosmological
solutions~\cite{Chamseddine:2011bu,Kobayashi:2012fz}, the black hole
solutions~\cite{Nieuwenhuizen:2011sq,Berezhiani:2011mt,Arraut:2013bqa,Kodama:2013rea}, the
Lema\^{i}tre-Tolman-Bondi (LTB) solution~\cite{Kobayashi:2012fz}, and the
Reissner-Nordstr\"{o}m (RN) solution~\cite{Berezhiani:2011mt}.
Since the equations of motion for
$g_{\mu\nu}$ (and, in fact, those for $f_{\mu\nu}$ as well) reduce to
the Einstein equations with a cosmological constant,
any spherically symmetric solution in GR is also a solution of
the one-parameter subclass (\ref{alpha3alpha4}) of bi-gravity and massive gravity with a
suitable fiducial metric.
In appendix~\ref{exampleofsolutions}, we
present some examples of bi-FLRW and bi-Schwarzschild-de Sitter
solutions belonging to this class.

\section{Linear perturbations}
\label{Sec.4}

Now, we analyze linear perturbations
around bi-spherically symmetric solutions
given in the previous section. The two tensor fields of metrics are perturbed as
\begin{eqnarray}
g_{\mu\nu}&=&\bar{g}_{\mu\nu} + \delta g_{\mu\nu},\\
f_{\mu\nu}&=&\bar{f}_{\mu\nu} + \delta f_{\mu\nu}.
\end{eqnarray}
The first-order perturbation, $\delta \gamma^{\mu}{}_{\nu}$, of $\gamma^{\mu}{}_{\nu}$ is defined as
\begin{eqnarray}
\sqrt{g^{-1}f} = \gamma^\mu{}_{\nu} = \bar{\gamma}^\mu{}_{\nu} + \delta\gamma^{\mu}{}_{\nu}+ 
{\cal O}(\text{second-order perturbations}),\label{deltagamma}
\end{eqnarray}
which can be written in terms of the metric perturbations by solving the
following equations,
\begin{eqnarray}
\bar{\gamma}^{\mu}{}_{\rho}\delta \gamma^{\rho}{}_{\nu}+\delta \gamma^{\mu}{}_{\rho} \bar{\gamma}^{\rho}{}_{\nu} &=& -\delta g^{\mu}{}_{\rho} \bar{\gamma}^{\rho}{}_{\sigma}\bar{\gamma}^{\sigma}{}_{\nu}+\delta f^{\mu}{}_{\nu},\label{dgamma=fg}
\end{eqnarray}
where $\delta g^{\mu}{}_{\nu} = \bar{g}^{\mu\rho}\delta g_{\rho\nu}$ and
$\delta f^{\mu}{}_{\nu} = \bar{g}^{\mu\rho}\delta f_{\rho\nu}$.
For our purpose we do not need the explicit form of the solution
to the above equation, though it is obtained for a general fiducial metric in ref.~\cite{Bernard:2014bfa}.
Actually, without the explicit form of $\delta\gamma^\mu{}_{\nu}$, we can
directly calculate $X^\mu{}_{\nu}$ from eq.~(\ref{deltagamma}) with
eq.~(\ref{gammabar}) as
\begin{eqnarray}
\delta X_{(g)}{}^\mu{}_{\nu}&=&C(t,r)
\begin{pmatrix}
0&0&0&0\\
0&0&0&0\\
0&0&\delta\gamma^{3}{}_{3}&-\delta\gamma{}^{2}{}_{3} \\
0&0&-\delta\gamma^{3}{}_{2}&\delta\gamma^{2}{}_{2} 
\end{pmatrix}.\label{deltaX}
\end{eqnarray}
Since the Einstein tensor satisfies the Bianchi identity
$\nabla^{(g)}_\mu G^{\mu}{}_{\nu}[g] = 0$ and the energy-momentum tensor
is conserved, the tensor $X_{(g)}{}^{\mu}{}_{\nu}$ also satisfies
$\nabla_{\mu} X_{(g)}{}^{\mu}{}_{\nu} =0$.  As demonstrated in
appendix~\ref{appBianchi}, this requirement leads to a stronger
condition
\begin{eqnarray}
 \delta X_{(g)}{}^{\mu}{}_{\nu} = 0\label{deltaX=0},
\end{eqnarray}
which yields $\delta \gamma^{a}{}_{b}=0$ for $a,b$ = $2,3$.
(Note that we are interested in the case with $C(t,r) \ne 0$.)
Since
\begin{eqnarray}
\delta X_{(f)}{}^\mu{}_{\nu}&=&-
\frac{1}{\kappa^2A^2|ad-bc|} \delta X_{(g)}{}^{\mu}{}_{\nu},\label{dXf}
\end{eqnarray}
eq.~(\ref{deltaX=0}) also implies $\delta X_{(f)}{}^\mu{}_\nu = 0$.  Thus, the equations of
motion for the linear perturbations $\delta g_{\mu\nu}$ and $\delta f_{\mu\nu}$
reduce to the linearized Einstein equations.

In order to see the implications of the equation (\ref{deltaX=0}) in
more detail, we express $\delta\gamma^\mu{}_\nu$ in terms of the metric perturbations.  Since
only the angular components of $\delta\gamma^\mu{}_\nu$
enter the equation (\ref{deltaX=0}),
we only have to deal with the angular components
of the equation (\ref{dgamma=fg}), which can easily be solved because
$\bar{\gamma}^{a}{}_{b}=A\delta^{a}{}_{b}$ for $a,b$ = $2,3$. In fact,
eq.~(\ref{dgamma=fg}) reduces to
\begin{eqnarray}
2A\delta\gamma^{a}{}_{b} =
-A^2 \delta g^{a}{}_{b}+\delta f^{a}{}_{b} = 0.
\end{eqnarray}

To sum up, the equations of motion for the first-order perturbations are
equivalent to the following three equations:
\begin{eqnarray}
&& \delta G[g]{}^{\mu}{}_{\nu}= \delta T_{(g)\,}{}^{\mu}{}_{\nu},\label{Gg}\\
&& \delta G[f]{}^{\mu}{}_{\nu} = \delta T_{(f)\,}{}^{\mu}{}_{\nu},\label{Gf}\\
&& A^2 \delta g_{ab}-\delta f_{ab} =0. \label{df=dg}
\end{eqnarray}
This is one of the main results of this investigation. The equations of motion
for the perturbations of the two metrics coincide with the perturbed Einstein
equations, though $\delta g_{\mu\nu}$ and $\delta f_{\mu\nu}$
 are subject to eq.~(\ref{df=dg}).

Then let us count the number of graviton degrees of freedom for this
perturbed system. Each symmetric tensor field of metric has ten components,
and there are, respectively, four constraints
(the Hamiltonian and momentum constraints) in eqs.~(\ref{Gg}) and~(\ref{Gf}), since
those equations are the same as the perturbed Einstein equations. Furthermore,
eq.~(\ref{df=dg}) gives three constraints among the angular components of the perturbed metrics.
We have four spacetime coordinates and hence
there are four gauge degrees of freedom representing the choice of coordinates.
In addition to those familiar gauge degrees of freedom, it turns out that
there still remains another gauge transformation retaining the equations of motion~(\ref{Gg}), (\ref{Gf}),
and~(\ref{df=dg}),
as explicitly shown in appendix~\ref{addgauge}.
Note that this gauge degree of freedom
corresponds to the ambiguity of the linear perturbations mentioned in
ref.~\cite{Kodama:2013rea} for the Schwarzschild-de Sitter solution in the dRGT theory.
Thus, the number of the remaining degrees of freedom is
$10\times2 - 4 \times 2 -3 - (4+1)= 4$, which coincides with that of two massless
gravitons.
We can confirm that
this is consistent with the result of the
Hamiltonian analysis given in appendix~\ref{hamiltonian}:
there are ten first class constraints and twelve second class constraints,
and hence there are 8 $(= 40 - 10 \times 2 - 12)$ degrees of freedom in phase space.

The above analysis can be applied to dRGT massive gravity only
with eqs. (\ref{Gg}) and (\ref{df=dg}) because the derivations of these equations do not
depend on the equation of motion for $f_{\mu\nu}$. Since, in this case,
$\delta f_{\mu\nu}$ is composed of St\"{u}ckelberg fields, the condition
(\ref{df=dg}) just determines perturbations of St\"{u}ckelberg fields. The
remaining variables $\delta g_{\mu\nu}$ are governed by the Einstein
equations and additional gauge symmetry appears as gauge degree of
freedom for St\"{u}ckelberg fields.

\section{Second-order perturbations}
 \label{Sec.5}

In the previous section, we have shown that the first-order
perturbations obey the perturbed Einstein equations and hence the
behavior of the perturbations coincides with that of GR, though
$\delta g_{\mu\nu}$ and $\delta f_{\mu\nu}$ are subject to
eq.~(\ref{df=dg}). One may then ask the question as to how one can
discriminate this class of solutions in bi-gravity from the
corresponding solutions in GR.

One possibility is to take into account the back reaction on the physical
metric $g_{\mu\nu}$ from $\delta f_{\mu\nu}$ at second order. For
this purpose, we incorporate second-order perturbations as follows:
\begin{eqnarray}
g_{\mu\nu} = \bar{g}_{\mu\nu}+\delta g_{\mu\nu} + g^{(2)}{}_{\mu\nu},\\
f_{\mu\nu} = \bar{f}_{\mu\nu}+\delta f_{\mu\nu} + f^{(2)}{}_{\mu\nu}.
\end{eqnarray}
The perturbed metrics now give rise to the second-order perturbations of
$\gamma^{\mu}{}_{\nu}$ as
\begin{eqnarray}
 \gamma^{\mu}{}_{\nu}
  = \bar{\gamma}^{\mu}{}_{\nu} + \delta \gamma^{\mu}{}_{\nu} + \gamma^{(2)}{}^{\mu}{}_{\nu} +
  {\cal O}(\text{third-order perturbations}),
\end{eqnarray}
where $\bar{\gamma}^{\mu}{}_{\nu}$ is the background quantity defined in eq.~(\ref{gammabar})
and $\delta \gamma^{\mu}{}_{\nu}$ satisfies eq.~(\ref{deltaX=0}), and hence
$\delta \gamma^{a}{}_{b} = 0$ for $a,b = 2,3$. The interaction term
in the equations of motion for $g_{\mu\nu}$ can be
calculated explicitly even at second order, and is given by
\begin{eqnarray}
 X_{(g)}{}^{(2)}{}^{\mu}{}_{\nu} = 
  \begin{pmatrix}
0&0&0&0\\
0&0&0&0\\
     0&0&X^{(2)}{}^{2}{}_{2}(t,r,\theta,\phi)& X^{(2)}{}^{2}{}_{3}(t,r,\theta,\phi)\\
     0&0&
     \frac{X^{(2)}{}^{2}{}_{3}(t,r,\theta,\phi)}{\sin^2\theta}
     &  X^{(2)}{}^{3}{}_{3}(t,r,\theta,\phi)
    \end{pmatrix}.
\end{eqnarray}
This tensor satisfies the conditions assumed in
appendix~\ref{appBianchi}, which, together with the Bianchi identity, yield
\begin{eqnarray}
 X_{(g)}{}^{(2)}{}^{\mu}{}_{\nu} = 0.
\end{eqnarray}
Even at second order, $X_{(g)}{}^{(2)}{}^{\mu}{}_{\nu}$ is
proportional to $X_{(f)}{}^{(2)}{}^{\mu}{}_{\nu}$,
\begin{eqnarray}
X_{(f)}{}^{(2)}{}^\mu{}_{\nu}&=&-
\frac{1}{\kappa^2A^2|ad-bc|} X_{(g)}{}^{(2)}{}^{\mu}{}_{\nu},\label{dXf2}
\end{eqnarray}
leading to $X_{(f)}{}^{(2)}{}^{\mu}{}_{\nu} = 0$ as well. These
conditions provide the relation between $g^{(2)}{}_{ab}$ and
$f^{(2)}{}_{ab}$ as follows:
\begin{eqnarray}
 \gamma^{(2)}{}^a{}_b = - \frac{1}{A^2-A(a+d)+ad- bc}
   \delta \gamma^a{}_{A}(\bar{\gamma}^{A}{}_{B}-(\bar{\gamma}^{C}{}_{C}-A)\delta^{A}{}_{B})\delta\gamma^{B}{}_{b},
\end{eqnarray}
for $a,b = 2, 3$ and $A,B,C = 0,1$.
Thus, the metric perturbations obey the perturbed Einstein
equations also at second order, and the number of graviton degrees of freedom coincides with
that of two massless gravitons even at second order.

This fact indicates that one cannot discriminate this class of solutions
from the corresponding solutions in GR even at second order,
unfortunately. On the other hand, this fact, fortunately, implies that
our solutions are free from non-linear instabilities even in cubic
action, which plague many cosmological solutions in massive gravity,
such as the diagonal open FLRW
solution~\cite{Gumrukcuoglu:2011ew,Gumrukcuoglu:2011zh,DeFelice:2012mx},
flat FLRW solution~\cite{D'Amico:2011jj,D'Amico:2012pi}, and de Sitter
solution~\cite{Koyama:2011xz,D'Amico:2012pi}.
 
\section{Conclusions and discussion}

In the present study we have investigated the perturbations of a class of
spherically symmetric solutions in massive gravity and
bi-gravity. First, we classified spherically symmetric solutions in
massive gravity and bi-gravity
and identified the specific class for which the background equations of motion
are identical to a set of the Einstein equations with a cosmological constant.
These solutions are allowed only
with the one-parameter family of $\alpha_3$ and $\alpha_4$
 satisfying eqs.~(\ref{alpha3alpha4}).
This class of solutions includes
many known solutions, e.g., the FLRW solutions in
ref.~\cite{Chamseddine:2011bu,Kobayashi:2012fz}, the Schwarzschild(-de
Sitter) solutions in
ref.~\cite{Nieuwenhuizen:2011sq,Berezhiani:2011mt,Arraut:2013bqa,Kodama:2013rea}, the
LTB solution in ref.~\cite{Kobayashi:2012fz}, and the RN solution in
ref.~\cite{Berezhiani:2011mt}. In fact, any spherically
symmetric solution in GR is included in this class with a
suitable choice of the fiducial metric $f_{\mu\nu}$.

Next, we have investigated linear perturbations on this class of
solutions. We have found that the interaction terms in the equations of
motion for both metrics, $\delta X_{(g)}{}^{\mu}{}_{\nu}$ and $\delta
X_{(f)}{}^{\mu}{}_{\nu}$, vanish thanks to the Bianchi identities, and
hence the equations of motion reduce to eqs.~(\ref{Gg})-(\ref{df=dg}),
which are the perturbed Einstein equations with the
relation~(\ref{df=dg}).

We have also found that, in addition to the usual gauge symmetry associated
with spacetime coordinate transformation, there is another
gauge symmetry of the linear perturbations given by eqs.~(\ref{gauge0})-(\ref{gauge3}),
which has already been known for the perturbations of the Schwarzschild de
Sitter solution in dRGT massive gravity~\cite{Kodama:2013rea}.

We have shown that the above result applies to second-order
perturbations as well. Thus, one cannot distinguish this class of
solutions in massive gravity and bi-gravity from the corresponding
solutions of GR up to second order. This fact, however, implies
that this class of solutions do not suffer from the non-linear
instabilities, which often appear in the other cosmological solutions in
massive gravity and bi-gravity. These aspects would suggest that
massive gravity or bi-gravity with this one parameter family in
$(\alpha_3, \alpha_4)$ may have additional fully non-linear symmetry,
which may be responsible for the stability. Further investigations are
necessary to clarify this point.

In this article, only spherically symmetric background solutions are
discussed. So, it is an interesting and open question whether the
results obtained in this article hold for more general background
solutions. Our analysis on the background solutions can at least be
applied to any $\bar{\gamma}$ having the form of eq.~(\ref{gammabar}) in
any basis vectors, because the background equations of
motion~(\ref{eomgbar}) can be obtained in an algebraic way from
eq.~(\ref{gammabar}) irrespective of a concrete expression for
$\bar{\gamma}$. Extending the above analysis to linear and non-linear
perturbations of more general solutions, however, is a non-trivial issue
simply because such analysis accompanies the derivatives. We will
address these issues in a future publication.

\acknowledgments

This work was in part supported by the JSPS Grant-in-Aid for Scientific
Research Nos. 24740161 (T.K.), 25287054 (M.Y.), 26610062 (M.Y.), the
JSPS Grant-in-Aid for Scientific Research on Innovative Areas
No. 15H05888 (T.K. and M.Y.), and the JSPS Research Fellowship for Young
Scientists, No. 26-11495 (D.Y.).

\appendix
\section{Concrete examples of background solutions}
\label{exampleofsolutions}

\subsection{Bi-cosmological solutions}

First we consider a family of bi-FLRW solutions, in which physical
metric takes the following FLRW form:
\begin{eqnarray}
 \bar{g}_{\mu\nu}dx^\mu dx^\nu = -dt^2
  + a^2(t)\left[\frac{dr^2}{1-K r^2} + r^2 d\Omega^2 \right].
\end{eqnarray}
Comparing this metric with eq.~(\ref{gbar}) yields $R(t,r) = a(t)r$. We
assume that $f_{\mu\nu}$ takes the same FLRW metric but in a coordinate
$(\tilde{t}(t,r),\tilde{r}(t,r),\theta,\phi)$ different from that of
$g_{\mu\nu}$,
\begin{eqnarray}
 f_{\mu\nu}dx^\mu dx^\nu &=&
  -d\tilde{t}^2 + b^2(\tilde{t})\left[\frac{d\tilde{r}^2}{1-\tilde{K}\tilde{r}^2}+\tilde{r}^2d\Omega^2\right]
  \notag\\
 &=& f_{00}dt^2 + 2 f_{01} dt dr + f_{11}dr^2 + b^2(\tilde{t}) \tilde{r}^2(t,r)d\Omega^2,
\end{eqnarray}
where
\begin{eqnarray}
 f_{00} &=& - \left(\frac{\partial \tilde{t}}{\partial t}\right)^2 + \frac{b^2(\tilde{t}(t,r))}{1-\tilde{K} \tilde{r}^2(t,r)}\left(\frac{\partial \tilde{r}}{\partial t}\right)^2,\\
 f_{01} &=& -\frac{\partial \tilde{t}}{\partial t}\frac{\partial \tilde{t}}{\partial r}
  + \frac{b^2(\tilde{t}(t,r))}{1-\tilde{K} \tilde{r}^2(t,r)}\frac{\partial \tilde{t}}{\partial t}\frac{\partial \tilde{t}}{\partial r},\\
 f_{11} &=& - \left(\frac{\partial \tilde{t}}{\partial r}\right)^2 + \frac{b^2(\tilde{t}(t,r))}{1-\tilde{K} \tilde{r}^2(t,r)}\left(\frac{\partial \tilde{r}}{\partial r}\right)^2.
\end{eqnarray}
In order to apply the results of the main body, the radial coordinate
$\tilde{r}$ is determined to satisfy the following relation,
\begin{eqnarray}
 \tilde{r}(t,r) &=& \frac{A R(t,r)}{b(\tilde t(t,r))}= A \frac{a(t)r}{b(\tilde{t}(t,r))},
\end{eqnarray}
while the time coordinate $\tilde{t}$ is arbitrary. In this case, the
equations of motion for both metrics become Einstein equations with
cosmological constants so that $a(t)$ and $b(\tilde{t})$ obey the
Friedmann equation with respect to each (cosmic) time, $t$ or
$\tilde{t}$. This kind of bi-FLRW solution becomes a slight
generalization of that found in ref.~\cite{Volkov:2011an}, in which a
specific choice of the coordinate $\tilde{t}$ is adopted.

For $b = 1$, $\tilde{K}=0$, and $\Lambda^{(f)}_{\rm eff}=0$, the fiducial metric
$f_{\mu\nu}$ becomes the flat Minkowski one and hence this
bi-cosmological solution includes that obtained in
ref.~\cite{Chamseddine:2011bu,Kobayashi:2012fz} in dRGT massive
gravity with the flat fiducial metric.

\subsection{Bi-Schwarzschild de Sitter solutions}

Our results are applied to the following bi-Schwarzschild de Sitter
metrics as well:
\begin{eqnarray}
 g_{\mu\nu}dx^{\mu} dx^{\nu}
  &=&
  -\left(1-\frac{r_{(g)}}{r} + \Lambda^{(g)}_{\rm eff} r^2 \right)dt^2 +\frac{dr^2}{1-\frac{r_{(g)}}{r}+\Lambda^{(g)}_{\rm eff}r^2} + r^2 d\Omega^2, \\
 f_{\mu\nu}dx^{\mu} dx^{\nu}
  &=&
  -\left(1-\frac{\tilde{r}_{(f)}}{\tilde{r}} + \Lambda^{(f)}_{\rm eff} \tilde{r}^2 \right)d\tilde{t}^2 +\frac{d\tilde{r}^2}{1-\frac{\tilde{r}_f}{\tilde{r}}+\Lambda^{(f)}_{\rm eff}\tilde{r}^2} + \tilde{r}^2 d\Omega^2
\end{eqnarray}
with
\begin{eqnarray}
 \tilde{r} = A r,
\end{eqnarray}
where $r_{(g)}$ and $\tilde{r}_{(f)}$ represent Schwarzschild radii, and
$\Lambda^{(g)}_{\rm eff}$ and $\Lambda^{(f)}_{\rm eff}$ are effective
cosmological constants defined in eqs.~(\ref{Lambdagbar}) and
(\ref{Xefff}), respectively. Since the Schwarzschild-de Sitter metric is
a solution of Einstein equation with cosmological constant, this is a
vacuum solution in our setting. As is the case with the cosmological
solution, this black hole solution can be obtained with arbitrary choice
of the time coordinate $\tilde{t}(t,r)$.  By tuning the parameters
$\beta_0$ and $\beta_4$, we can set $\Lambda_{\rm eff}^{(g)}$ and
$\Lambda^{(f)}_{\rm eff}$ to be zeros simultaneously, which
corresponds to a bi-Schwarzschild solution.

\section{Bianchi identity}
\label{appBianchi}

In this appendix, we will show that a symmetric tensor satisfying a
condition given below must vanish as long as it obeys Bianchi identity
and the background metric $\bar{g}_{\mu\nu}$ takes the matrix form of eq.~(\ref{gbar}).

Let us consider the following symmetric tensor $X_{\mu\nu}$:
\begin{eqnarray}
 X^{\mu}{}_{\nu} =
  \Lambda \delta^{\mu}_{\nu}
  + \epsilon^n X^{(n)}{}^\mu{}_{\nu}
  +{\cal O}(\epsilon^{n+1})\label{X=L+Xn}
\end{eqnarray}
with
\begin{eqnarray}
 X^{(n)}{}^0{}_{\mu} =  X^{(n)}{}^1{}_{\mu} = X^{(n)}{}^{\mu}{}_{0} = X^{(n)}{}^{\mu}{}_{1} = 0,
\end{eqnarray}
where $\epsilon$ denotes the order of perturbations and $\Lambda$ is a
constant. The goal of this section is to show that
$X^{(n)}{}^{\mu}{}_{\nu}$ vanishes if the Bianchi identity,
$\nabla_{\mu} X^{\mu}{}_{\nu} = 0$, is imposed.

The tensor $\bar{g}_{\mu\rho}X^{(n)}{}^{\rho}{}_{\nu}$ is symmetric because
\begin{eqnarray}
 X_{\mu\nu}
  &=& g_{\mu\rho}X^{\rho}{}_{\nu}\\
 &=& \Lambda g_{\mu\nu}
  +\bar{g}_{\mu\rho} X^{(n)}{}^{\rho}{}_{\nu} +{\cal O}(\epsilon^{n+1}),
\end{eqnarray}
and both of $X_{\mu\nu}$ and $\Lambda g_{\mu\nu}$ are symmetric. Then,
from the property of the background metric $\bar{g}_{\mu\nu}$, it is
characterized by three arbitrary functions as follows:
\begin{eqnarray}
 \bar{g}_{\mu\rho} X^{(n)}{}^{\rho}{}_{\nu} =
  \begin{pmatrix}
   0&0&0&0\\
   0&0 &0 &0 \\
   0&0 &X^{(n)}_{22}(t,r,\theta,\phi)  & X^{(n)}_{23}(t,r,\theta,\phi)\\
   0&0&X^{(n)}_{23}(t,r,\theta,\phi) & X^{(n)}_{33}(t,r,\theta,\phi)
  \end{pmatrix},\label{generalXmn}
\end{eqnarray}
or equivalently,
\begin{eqnarray}
  X^{(n)}{}^{\mu}{}_{\nu} =
  \begin{pmatrix}
   0&0&0&0\\
   0&0 &0 &0 \\
   0&0 &\frac{X^{(n)}_{22}(t,r,\theta,\phi)}{R^2(t,r)}  & \frac{X^{(n)}_{23}(t,r,\theta,\phi)}{R^2(t,r)}\\
   0&0&\frac{X^{(n)}_{23}(t,r,\theta,\phi)}{R^2(t,r)\sin^2\theta} & \frac{X^{(n)}_{33}(t,r,\theta,\phi)}{R^2(t,r)\sin^2\theta}
  \end{pmatrix}.\label{genXn}
\end{eqnarray}
On the other hand, from the eq.~(\ref{X=L+Xn}), the Bianchi identity reads
\begin{eqnarray}
 \nabla_{\mu} X^\mu{}_{\nu} = \epsilon^n \bar{\nabla}_{\mu} X^{(n)}{}^\mu{}_{\nu} +{\cal O}(\epsilon^{n+1})=0,
\end{eqnarray}
where $\bar{\nabla}_{\mu}$ is the covariant derivative with respect to
$\bar{g}_{\mu\nu}$. Then, zero-th and first components of this equation
are
\begin{eqnarray}
 \nabla_\mu X^{\mu}{}_{0} &=& -(X^{(n)}_{22} + (\sin\theta)^{-2} X^{(n)}_{33}) \frac{\partial_t R}{R^3} \epsilon^n + {\cal O}(\epsilon^{n+1}) =0,\\
 \nabla_\mu X^{\mu}{}_{1} &=& -(X^{(n)}_{22} + (\sin\theta)^{-2} X^{(n)}_{33}) \frac{\partial_r R}{R^3}
  \epsilon^n + {\cal O}(\epsilon^{n+1})
  =0,
\end{eqnarray}
which yields the following solution when $R$ is not a constant,
\begin{eqnarray}
 X^{(n)}_{22}(t,r,\theta,\phi) = -\frac{X^{(n)}_{33}(t,r,\theta,\phi)}{\sin^2 \theta}.\label{X22=}
\end{eqnarray}
The remaining components of this equation are given by
\begin{eqnarray}
 \nabla_\mu X^{\mu}{}_{2} &=& (R \sin\theta)^{-2}\left( \partial_{\phi}X^{(n)}_{23}- \partial_{\theta}X^{(n)}_{33}\right) \epsilon^n + {\cal O}(\epsilon^{n+1})=0,\label{bianchitheta}\\
 \nabla_\mu X^{\mu}{}_{3} &=& (R \sin\theta)^{-2}\left( \partial_{\phi}X^{(n)}_{33}+\sin\theta \partial_{\theta}(\sin\theta X^{(n)}_{23})\right) \epsilon^n + {\cal O}(\epsilon^{n+1})=0,\label{bianchiphi}
\end{eqnarray}
where we have used the relation~(\ref{X22=}). Removing $X^{(n)}_{23}$
from these equations leads to the following equation for $X^{(n)}_{33}$:
\begin{eqnarray}
 \frac{1}{\sin\theta}\partial_{\theta}(\sin\theta \partial_\theta X^{(n)}_{33}) + \frac{1}{\sin\theta^2}\partial_{\phi}\partial_{\phi} X^{(n)}_{33} = 0.\label{harmonic}
\end{eqnarray}
Since this is just the Laplace equation on a sphere, its solution is
constant over the sphere:
\begin{eqnarray}
 X^{(n)}_{33} = f(t,r).
\end{eqnarray}
By plugging this solution into eqs. (\ref{bianchitheta}) and (\ref{bianchiphi}), we obtain 
\begin{eqnarray}
 X^{(n)}_{23} &=& \frac{g(t,r)}{\sin \theta}.
\end{eqnarray}
Thus, the solution of the Bianchi identity is given by
\begin{eqnarray}
 X^{(n)}_{\mu\nu}
  =
\begin{pmatrix}
   0&0&0&0\\
   0&0 &0 &0 \\
   0&0 &-\frac{f(t,r)}{\sin^2 \theta}  & \frac{g(t,r)}{\sin \theta}\\
   0&0 & \frac{g(t,r)}{\sin \theta}  & f(t,r)
  \end{pmatrix}  .
\end{eqnarray}
However, the components with $\sin \theta$ in their denominators are singular at
$\theta = 0, \pi$ unless
\begin{eqnarray}
 f(t,r) = 0,\\
g(t,r) = 0.
\end{eqnarray}
Therefore, the regular solution of $\nabla_\mu X^{\mu}{}_{\nu}=0$ is
\begin{eqnarray}
 X^{(n)}_{\mu\nu} = 0.
\end{eqnarray}

\section{Additional gauge symmetry of linear perturbations}
\label{addgauge}

The linear perturbations have an additional gauge symmetry,
which is combination of gauge  transformation of $g_{\mu\nu}$ and $f_{\mu\nu}$ separately but keeping the equation~(\ref{df=dg}).
In
this appendix, we will give a concrete form of such coordinate
transformation.

For this purpose, let us consider
infinitesimal gauge transformation generated by $x^\mu\rightarrow x^{\mu} - \xi^\mu$ for $g_{\mu\nu}$ and
$x^\mu\rightarrow x^{\mu} - (\xi^\mu+ \delta \xi^\mu)$ for $f_{\mu\nu}$\footnote{To determine the gauge transformation, one establish a bi-tangent bundle $T^2M$ i.e. a fibre bundle locally isomorphic to $M\times T^{(g)}_p\times T^{(f)}_p$. To have a usual tangent bundle $TM$, two horizontal lifts $\pi_{(g)}^{-1}(M)$ and $\pi_{(f)}^{-1}(M)$ are identified by this relation of the diffeomorphisms so that it determines a diffeomorphism group of the base manifold.}.

We denote the difference $A^2 \delta g_{ab}-\delta f_{ab}$ in
eq.~(\ref{df=dg}) under this transformation by $\Delta_{ab}$, that is,
\begin{eqnarray}
 A^2 \delta g_{ab}-\delta f_{ab}
  \rightarrow  A^2 \delta g_{ab}-\delta f_{ab}
  + \Delta_{ab}.
\end{eqnarray}
The additional gauge symmetry is characterized by $\Delta_{ab}=0$.  The
$(2,2)$ component of this condition is given by
\begin{eqnarray}
0 = \frac{\Delta_{22}}{-2A^2R} = \delta \xi^0 \partial_t R+  \delta \xi^1 \partial_r R + R \partial_{\theta} \delta \xi^2.\label{eq22c}
\end{eqnarray}
The remaining $(2,3)$ and $(3,3)$ components are given by
\begin{eqnarray}
 0 &=& \frac{\Delta_{23}}{-A^2 R^2 \sin \theta} =
  \partial_{\phi} \left(\frac{\delta \xi^2}{\sin \theta}\right) + \sin \theta \partial_{\theta} \delta\xi^3,\label{D23} \\
 0&=& \frac{\Delta_{33}}{-2 A^2 R^2 \sin^3\theta}
  = -\partial_{\theta}\left( \frac{\delta\xi^2}{\sin \theta}\right) + \frac{ \partial_{\phi} \delta\xi^3}{\sin\theta},\label{D33}
\end{eqnarray}
where we have used eq.~(\ref{eq22c}). One can easily find, similarly to
eq.~(\ref{harmonic}), that these equations reduce to the Laplace
equation on a sphere:
\begin{eqnarray}
\frac{1}{\sin \theta} \partial_{\theta}\left(\sin\theta \partial_{\theta} \delta \xi^3 \right) + \frac{1}{\sin^2\theta}\partial_{\phi}\partial_{\phi} \delta \xi^3 = 0,
\end{eqnarray}
whose solution becomes
\begin{eqnarray}
 \delta \xi^3 = P(t,r).
\end{eqnarray}
Plugging this solution into eqs.~(\ref{D23}) and (\ref{D33}) we find
\begin{eqnarray}
  \delta \xi^2 = Q(t,r) \sin \theta.
\end{eqnarray}
To sum up, this additional gauge symmetry is characterized by
$\Xi(t,r,\theta,\phi) , P(t,r), Q(t,r)$ as
\begin{eqnarray}
  \delta \xi^0 &=& \Xi(t,r,\theta,\phi),\label{gauge0} \\
  \delta \xi^1 &=& -\frac{\partial_{t} R(t,r) \Xi(t,r,\theta,\phi)+R(t,r)Q(t,r) \cos \theta}{\partial_{r}R(t,r)},\label{gauge1} \\
  \delta \xi^2 &=& Q(t,r)\sin\theta,\label{gauge2}\\
  \delta \xi^3 &=& P(t,r).\label{gauge3}
\end{eqnarray}
One may regard $R(t,r)$ itself as a radial coordinate and, in the new
coordinates $(t,R,\theta,\phi)$, the above transformation
(\ref{gauge0})-(\ref{gauge3}) with $P(t,r)=Q(t,r)=0$ simply reduces
to the transformation of the time coordinate.
 
We can directly observe this symmetry in the action. Actually, the quadratic
action of the mass term for the linear perturbations becomes
\begin{eqnarray}
  S^{(2)}_{\rm mass} &=& \frac{M_{\rm pl}^2}{2}
   \int d^4 x \sqrt{-\bar{g}}\frac{2 C (t,r)}{A R^2} (\delta \gamma^{2}{}_{2}\delta\gamma^{3}{}_{3}
   - \delta \gamma^{2}{}_{3}\delta\gamma^{3}{}_{2}) \notag\\
  &&+
   \frac{M_{\rm pl}^2}{2}
   \int d^4 x \sqrt{-g}^{(2)}(-2 \Lambda_{\rm eff}^{(g)})
    +
   \frac{\kappa^2 M_{\rm pl}^2}{2}
   \int d^4 x \sqrt{-f}^{(2)}(-2 \Lambda_{\rm eff}^{(f)}),\label{Lmass}
\end{eqnarray}
where $C(t,r)$ is the function defined in eq.~(\ref{Cr}),
$\Lambda^{(g)}_{\rm eff}$ and $\Lambda^{(f)}_{\rm eff}$ are effective
cosmological constants defined in eqs.~(\ref{Lambdagbar}) and
(\ref{Xefff}), $\sqrt{-g}^{(2)}$ and $\sqrt{-f}^{(2)}$ are quadratic
perturbations of $\sqrt{-g}$ and $\sqrt{-f}$.  Clearly, this term is
invariant under the transformation~(\ref{gauge0})-(\ref{gauge3}) because
this transformation leaves $\delta \gamma^{a}{}_{b}$ unchanged.

\section{Hamiltonian analysis of linear perturbations}
\label{hamiltonian}

We will count the number of graviton degrees of freedom of linear
perturbations by means of the Hamiltonian analysis. So, we omit the
matter action in this appendix. For this purpose, it is useful to
decompose the perturbations in terms of spherical harmonics $Y^m_{l}$ as
done in ref.~\cite{Regge:1957td}. Due to the spherical symmetry of the
background metrics, the modes with different eigenvalues of rotation
($l,m$) or parity (odd or even) develop independently, and the dynamics
of each mode does not depend on $m$. Hence, we may suppose that $m$ is
equal to zero, without loss of generality.

\subsection{Odd mode perturbations}

Non-vanishing components of the odd mode perturbations with $m=0$
are given by
\begin{align}
 &\delta g_{03} = \sum_{l\geq 1} h^{(g),l}_{0}(t,r) \sin \theta \partial_{\theta}P_{l}(\cos \theta),\\
 &\delta g_{13} = \sum_{l\geq 1} h^{(g),l}_{1}(t,r) \sin \theta \partial_{\theta}P_{l}(\cos \theta),\\
 &\delta g_{23} = \sum_{l\geq 2} h^{(g),l}_{2}(t,r) \sin^2 \theta \partial_{\theta}\left(\frac{\partial_{\theta}P_{l}(\cos \theta)}{\sin \theta} \right),
\end{align}
and
\begin{align}
 &\delta f_{03} = \sum_{l\geq 1} h^{(f),l}_{0}(t,r) \sin \theta \partial_{\theta}P_{l}(\cos \theta),\\
 &\delta f_{13} = \sum_{l\geq 1} h^{(f),l}_{1}(t,r) \sin \theta \partial_{\theta}P_{l}(\cos \theta),\\
 &\delta f_{23} = \sum_{l\geq 2} h^{(f),l}_{2}(t,r) \sin^2 \theta \partial_{\theta}\left(\frac{\partial_{\theta}P_{l}(\cos \theta)}{\sin \theta} \right),
\end{align}
where $P_l$ is the Legendre polynomial. In this subsection, hereafter,
we omit the suffix $l$ and the summation with respect to $l$ for
brevity. From the perturbed Einstein-Hilbert action with the mass term
(\ref{Lmass}), the conjugate momenta of $h_I^{(g/f)}~(I=0,1,2)$ are
calculated as
\begin{eqnarray}
 P_0^{(g)} &=& \frac{\delta S}{\delta \dot{h}_0^{(g)}}= 0,\\
 P_1^{(g)} &=& \frac{\delta S}{\delta \dot{h}_1^{(g)}}= \frac{2\tilde{M}_{\rm pl}^2}{\sqrt{-\bar{g}}}\left( 2  (\ln R)' h_0^{(g)} -h_0^{(g)}{}' + \dot{h}_1^{(g)}\right),\\
 P_2^{(g)} &=& \frac{\delta S}{\delta \dot{h}_2^{(g)}}= \frac{2 \lambda \tilde{M}_{\rm pl}^2}{R^2} \sqrt{-\bar{g}}  \left(\bar{g}^{00} h^{(g)}_0 + \bar{g}^{01} h^{(g)}_1 - \bar{g}^{00}\dot{h}^{(g)}_2 - \bar{g}^{01}h^{(g)}_2{}' \right),\\
  P_0^{(f)} &=& \frac{\delta S}{\delta \dot{h}_0^{(l)}}= 0,\\
 P_1^{(f)} &=& \frac{\delta S}{\delta \dot{h}_1^{(l)}} =\frac{2\kappa^2\tilde{M}_{\rm pl}^2}{\sqrt{-\bar{f}}}\left( 2  (\ln R)' h_0^{(f)} -h_0^{(f)}{}' + \dot{h}_1^{(f)}\right),\\
 P_2^{(f)} &=&\frac{\delta S}{\delta \dot{h}_2^{(l)}}= \frac{2 \lambda \kappa^2 \tilde{M}_{\rm pl}^2}{A^2 R^2} \sqrt{-\bar{f}}  \left(\bar{f}^{00} h^{(f)}_0 + \bar{f}^{01} h^{(f)}_1 - \bar{f}^{00}\dot{h}^{(f)}_2 - \bar{f}^{01}h^{(f)}_2{}' \right),
\end{eqnarray}
where $\lambda := (l-1)(l+2)$, $\tilde{M}_{\rm pl}^2 := \frac{l (1+l)
}{1+2l}M_{\rm pl}^2 \pi$, and $\sqrt{-\bar{g}}$ represents the
determinant of only $0,1$ components:
\begin{eqnarray}
 \sqrt{-\bar{g}} := \sqrt{-\bar{g}_{00}\bar{g}_{11}+(\bar{g}_{01})^2 }.
\end{eqnarray}

We schematically decompose the Hamiltonian density as follows:
\begin{eqnarray}
 {\cal H}^{\rm odd} &=& {\cal H}^{\rm odd}_{GR,(g)} +{\cal H}^{\rm odd}_{GR,(f)}+{\cal H}^{\rm odd}_{\rm mass},
\end{eqnarray}
where ${\cal H}^{\rm odd}_{GR,(g/f)}$ represents the contribution from
each Einstein-Hilbert term and the effective cosmological term, which is
the second term (or third term) in the right hand side of
eq.~(\ref{Lmass}).  ${\cal H}^{\rm odd}_{\rm mass}$ represents the
contribution from the first term in the right hand side of
eq.~(\ref{Lmass}).  This decomposition is justified because $S_{mass}$
does not include time derivative of $g_{\mu\nu}$ and $f_{\mu\nu}$.  From
the expression of the action (\ref{Lmass}), ${\cal H}^{\rm odd}_{\rm
mass}$ is explicitly calculated as
\begin{eqnarray}
 {\cal H}^{\rm odd}_{\rm mass} = \tilde{M}_{\rm pl}^2 \lambda C(t,r)\frac{\sqrt{-\bar{g}}}{A^3 R^4} \left(A^2 h_2^{(g)} -h_2^{(f)}\right)^2.
\end{eqnarray}
It should be noted that, for $l=1$ mode, ${\cal H}^{\rm odd}_{\rm mass}$
vanishes, which implies that dynamics of $l=1$ mode coincides with that
of GR. Therefore, there should be additional gauge symmetry, under which
each metric transforms independently. This transformation, actually,
corresponds to the arbitrary function $P(t,r)$ in eq.~(\ref{gauge3}).

From now on, we focus on $l \geq 2$ modes.  The Hamiltonian density from the
Einstein-Hilbert term is calculated as
\begin{eqnarray}
  {\cal H}^{\rm odd}_{GR,(g)} 
  &=&
  \frac{1}{4 \tilde{M}_{\rm pl}^2} \sqrt{-\bar{g}} (P^{(g)}_1)^2
  + \frac{1}{4 \tilde{M}_{\rm pl}^2 \lambda} \frac{\sqrt{-\bar{g}} R^2}{\bar{g}_{11}} (P^{(g)}_2)^2
  + \frac{\bar{g}_{01}}{\bar{g}_{11}}(-h^{(g)}_1 + h^{(g)}_2{}') P^{(g)}_2\notag\\
  &&
   +\tilde{M}_{\rm pl}^2 \lambda \frac{\sqrt{-\bar{g}}}{\bar{g}_{11}R^2} (h^{(g)}_2{}'){}^2
   -2\tilde{M}_{\rm pl}^2 \lambda \frac{\sqrt{-\bar{g}}}{\bar{g}_{11}R^2} h^{(g)}_1 h^{(g)}_2{}'
   + 4 \tilde{M}_{\rm pl}^2 \lambda \sqrt{-\bar{g}} \frac{\bar{g}^{1A}\partial_A (\ln R)'}{R^2}h^{(g)}_1 h^{(g)}_2
   \notag\\
  &&
   +M^{(g)}_1 (h^{(g)}_1){}^2 + M^{(g)}_2 (h^{(g)}_2){}^2 - h^{(g)}_0 {\cal C}^{(1)}_{(g)}[P^{(g)}_1,P^{(g)}_2,h^{(g)}_1,h^{(g)}_2],
\end{eqnarray}
where $M^{(g)}_1, M^{(g)}_2$ are some functions of $t,r$ and ${\cal C}^{(1)}_{(g)}$
is given by
\begin{eqnarray}
 {\cal C}^{(1)}_{(g)} &=& 2  (\ln R)' P^{(g)}_1 + P^{(g)}_1{}' - P^{(g)}_2
  - 4 \tilde{M}_{\rm pl}^2 \frac{(\ln R)\dot{}}{\sqrt{-\bar{g}}} h^{(g)}_1{}'\notag\\
&&  + 4 \lambda \tilde{M}_{\rm pl}^2 \sqrt{-\bar{g}} \bar{g}^{0A}\partial_A(\ln R)\frac{1}{R^2} h^{(g)}_2
  - 2 \tilde{M}_{\rm pl}^2 \frac{1}{R^2}\partial_r \left(\frac{(R^2)\dot{}}{\sqrt{-\bar{g}}}\right) h^{(g)}_1,
\end{eqnarray}
with $A = 0,1$.
${\cal H}^{\rm odd}_{GR,(f)}$ is obtained by replacing $g \rightarrow f$,
$R \rightarrow A R$, $\tilde{M}_{\rm pl}^2 \rightarrow \kappa^2
\tilde{M}_{\rm pl}^2$.

The primary constraints of this system are
\begin{eqnarray}
 {\cal C}_{(g)}^{(0)} &:=& P^{(g)}_0 \approx 0,\\
 {\cal C}_{(f)}^{(0)} &:=& P^{(f)}_0 \approx 0,
\end{eqnarray}
and then, the total Hamiltonian is
\begin{eqnarray}
 H^{\rm odd}_T &=& H^{\rm odd} + \int d r \left[v^{(g)}(t,r) {\cal C}_{(g)}^{(0)}+v^{(f)}(t,r) {\cal C}_{(f)}^{(0)}\right], \\
 H^{\rm odd} &=& H^{\rm odd}_{GR,(g)} + H^{\rm odd}_{GR,(f)} + H^{\rm odd}_{\rm mass},\\
  H^{\rm odd}_{GR,(g/f)} &=& \int d r {\cal H}^{\rm odd}_{GR,(g/f)},
  \quad H^{\rm odd}_{\rm mass} = \int d r {\cal H}^{\rm odd}_{\rm mass}.
\end{eqnarray}
Time evolution of the primary constraints is given by
\begin{eqnarray}
 \dot{{\cal C}}_{(g/f)}^{(0)} =\{{\cal C}_{(g/f)}^{(0)}, H_T^{\rm odd}\} \approx 
  {\cal C}_{(g/f)}^{(1)}[P_1^{(g/f)},P_2^{(g/f)},h_1^{(g/f)},h_2^{(g/f)}],
\end{eqnarray}
which generate the following two secondary constraints,
\begin{eqnarray}
 {\cal C}_{(g/f)}^{(1)} \approx 0.
\end{eqnarray}
Time evolution of ${\cal C}_{(g)}^{(1)}$ is given by
\begin{eqnarray}
 \dot{{\cal C}}_{(g)}^{(1)} &=& \frac{\partial {\cal C}_{(g)}^{(1)}}{\partial t} + \{{\cal C}_{(g)}^{(1)},H^{\rm odd}_T\} \approx -\{P_2^{(g)} , H^{\rm odd}_{\rm mass}\}
  \notag\\
  &=&
  2 \lambda \tilde{M}_{\rm pl}^2\frac{\sqrt{-\bar{g}}}{A R^4}C(t,r) (A^2 h^{(g)}_2-h^{(f)}_2),
\end{eqnarray}
and that of ${\cal C}_{(f)}^{(1)}$ is given by
\begin{eqnarray}
 \dot{{\cal C}}_{(f)}^{(1)} 
  &\approx&
  - 2 \lambda \tilde{M}_{\rm pl}^2\frac{\sqrt{-\bar{g}}}{A^3 R^4}C(t,r) (A^2 h^{(g)}_2-h^{(f)}_2) .
\end{eqnarray}
These equations impose another constraint,
\begin{eqnarray}
 {\cal C}^{(2)} := A^2 h^{(g)}_2 - h_2^{(f)} \approx 0.
\end{eqnarray}
From time evolution of ${\cal C}^{(2)}$,
\begin{eqnarray}
 \dot{{\cal C}}^{(2)} &=& \{{\cal C}^{(2)}, H^{\rm odd}_T \}\notag\\
 &\approx&
  A^2 \{h_2^{(g)} , H^{\rm odd}_{GR,(g)}\}-\{h_2^{(f)} , H^{\rm odd}_{GR,(f)}\}\notag\\
 &\approx& A^2 \left(h^{(g)}_0
  - \frac{\bar{g}_{01}}{\bar{g}_{11}} h^{(g)}_1
  + \frac{\sqrt{\bar{g}}R^2}{2\lambda \tilde{M}_{\rm pl}^2 \bar{g}_{11}} P^{(g)}_2 \right)\notag\\&&
 -\left(h^{(f)}_0
  - \frac{\bar{f}_{01}}{\bar{f}_{11}} h^{(f)}_1
  + \frac{\sqrt{\bar{f}}A^2 R^2}{2 \lambda \kappa^2 \tilde{M}_{\rm pl}^2 \bar{f}_{11}} P^{(f)}_2\right),
\end{eqnarray}
we obtain yet another constraint,
\begin{eqnarray}
 {\cal C}^{(3)} &:=& 
  A^2 \left(h^{(g)}_0
  - \frac{\bar{g}_{01}}{\bar{g}_{11}} h^{(g)}_1
  + \frac{\sqrt{\bar{g}}R^2}{2\lambda \tilde{M}_{\rm pl}^2 \bar{g}_{11}} P^{(g)}_2 \right)\notag\\&&
 -\left(h^{(f)}_0
  - \frac{\bar{f}_{01}}{\bar{f}_{11}} h^{(f)}_1
  + \frac{\sqrt{\bar{f}}A^2 R^2}{2 \lambda \kappa^2 \tilde{M}_{\rm pl}^2 \bar{f}_{11}} P^{(f)}_2\right)\label{C3}.
\end{eqnarray}
Since ${\cal C}^{(3)}$ includes $h_0^{(g)}$ and $h_0^{(f)}$ terms, the Poisson
brackets of ${\cal C}^{(3)}$ and primary constraints ${\cal
C}^{(0)}_{(g/f)}$ do not vanish. Thus, the consistency relation on
${\cal C}^{(3)}$,
\begin{eqnarray}
 \dot{{\cal C}}^{(3)} 
  \approx \partial_t {\cal C}^{(3)}+  \{{\cal C}^{(3)},H\} + A^2 v_{(g)}-v_{(f)} \approx 0,
\end{eqnarray}
determines the combination of multipliers,
$A^2 v_{(g)} -v_{(f)}$. Then, no further constraints are generated.

Since one multiplier remains undetermined, one can easily find that there is gauge
symmetry in this system. More explicitly, one can confirm that there are
two first class constraints (and four second class constraints) in this
system through the presence of two zero eigenvalues of 6 $\times$ 6
matrix $\{{\cal C}_I , {\cal C}_J\}$, where ${\cal C}_I$ represent all
of the six constraints.
These two gauge symmetries correspond to the ones which the theory originally possesses.
To summarize, the number of graviton degrees of
freedom in this system is
\begin{eqnarray}
 \frac{1}{2}
 \left( \overbrace{12}^{\text{variables}} - \overbrace{6}^{\text{constraints}} 
 - \overbrace{2}^{\text{gauge dofs} } \right) = 2,
\end{eqnarray}
and completely coincides with the case of two massless gravitons.

For the
$l=1$ mode, there are four variables (eight variables in phase space), $h_0{}^{(g/f)},h_1{}^{(g/f)}$. As mentioned above, the interaction term $S_{mass}$ vanishes for $l=1$ mode, and hence the action reduces to decoupled two Einstein-Hilbert action.
Then, there are four first class constraints and four gauge symmetries which correspond to the general covariance of $g_{\mu\nu}$ and $f_{\mu\nu}$ separately. These four gauge symmetries can be arranged into the ones of the full theory and the additional ones described by $P(t,r)$ in~eq.(\ref{gauge3}).
To summarize, the number of degrees of freedom of the odd $l=1$ mode is
\begin{eqnarray}
 \frac{1}{2}
 \left( \overbrace{8}^{\text{variables}} - \overbrace{4}^{\text{constraints}} 
 - \overbrace{4}^{\text{gauge dofs} } \right) = 0.
\end{eqnarray}

\subsection{Even mode perturbations}

Similarly we consider even mode perturbations. Non-vanishing components
of the even mode perturbations are given by
\begin{eqnarray}
  \delta g_{00} &=& \sum_{l \geq 0}H^{(g),l}_0(t,r) P_l(\cos \theta), \\
  \delta g_{01} &=& \sum_{l \geq 0}H^{(g),l}_1(t,r) P_l(\cos \theta), \\
  \delta g_{02} &=& \sum_{l \geq 1}H^{(g),l}_2(t,r) \partial_{\theta}P_l(\cos \theta),\\
  \delta g_{11} &=& \sum_{l\geq 0}H^{(g),l}_3(t,r) P_{l}(\cos\theta),\\
  \delta g_{12} &=& \sum_{l\geq 1}H^{(g),l}_4(t,r) \partial_{\theta}P_{l}(\cos\theta),\\
  \delta g_{22} &=& \sum_{l \geq 0}H^{(g),l}_5(t,r) P_{l}(\cos\theta) + \sum_{l \geq 2}H^{(g),l}_6 \partial_{\theta}\partial_{\theta}P_{l}(\cos\theta),\\
  \delta g_{33} &=& \sum_{l \geq 0}H^{(g),l}_5(t,r) \sin^2 \theta P_{l}(\cos\theta) + \sum_{l \geq 2} H_6^{(g),l} \sin\theta \cos\theta \partial_{\theta}P_{l}(\cos\theta),
\end{eqnarray}
and similar expansions are applied for $\delta f_{\mu\nu}$. In this
subsection, hereafter, we omit the suffix $l$ and the summation with
respect to $l$ for brevity. First we treat the $l\geq 2$ modes, and the
Hamiltonian density for even modes is decomposed into
\begin{eqnarray}
  {\cal H}^{\rm even} = {\cal H}^{\rm even}_{GR,(g)}+ {\cal H}^{\rm even}_{GR,(f)} + {\cal H}^{\rm even}_{\rm mass}.
\end{eqnarray}
${\cal H}^{\rm even}_{GR,(g/f)}$ represents a contribution from
the Einstein-Hilbert term and effective cosmological constant terms,
explicitly given by
\begin{eqnarray}
  {\cal H}^{\rm even}_{GR,(g/f)} &=&
   - H^{(g/f)}_0 {\cal C}^{(1)}_{0,(g/f)}[P^{(g/f)}_3,P^{(g/f)}_5,H^{(g/f)}_3,H^{(g/f)}_4,H^{(g/f)}_5,H^{(g/f)}_6] \notag\\
  &&- H^{(g/f)}_1 {\cal C}^{(1)}_{1,(g/f)}[P^{(g/f)}_3,P^{(g/f)}_4,P^{(g/f)}_5,H^{(g/f)}_3,H^{(g/f)}_4,H^{(g/f)}_5,H^{(g/f)}_6] \notag\\
&&  - H^{(g/f)}_2 {\cal C}^{(1)}_{2,(g/f)}[P^{(g/f)}_4,P^{(g/f)}_6,H^{(g/f)}_3,H^{(g/f)}_4,H^{(g/f)}_5,H^{(g/f)}_6]  \notag\\
 && + \left(\text{second order terms of $H^{(g/f)}_3,H^{(g/f)}_4,H^{(g/f)}_5,H^{(g/f)}_6,P^{(g/f)}_3,P^{(g/f)}_4,P^{(g/f)}_5,P^{(g/f)}_6$} \right)
  \notag\\
\end{eqnarray}
with
\begin{eqnarray}
 {\cal C}^{(1)}_{0,(g)} &=& -R \bar{g}^{0I}(\partial_I R )P^{(g)}_5 + \left(\text{linear terms of }P^{(g)}_3,H^{(g)}_3,H^{(g)}_4,H^{(g)}_5,H^{(g)}_6 \right),\\
 {\cal C}^{(1)}_{1,(g)} &=& -2 R \bar{g}^{1I}(\partial_I R )P^{(g)}_5 + \left(\text{linear terms of }P^{(g)}_3,P^{(g)}_4,H^{(g)}_3,H^{(g)}_4,H^{(g)}_5,H^{(g)}_6 \right),\\
 {\cal C}^{(1)}_{2,(g)} &=& -2 P^{(g)}_6 +
  \left(\text{linear terms of } P^{(g)}_4,H^{(g)}_3,H^{(g)}_4,H^{(g)}_5,H^{(g)}_6 \right),
\end{eqnarray}
where $P^{(g/f)}_{I}$ are the conjugate momenta of $H^{(g/f)}_{I}$.
On the other hand, ${\cal H}^{\rm even}_{\rm mass}$ is given by
\begin{align}
  {\cal H}^{\rm even}_{\rm mass} = -\hat{M}_{\rm pl}^2 \frac{ \sqrt{-\bar{g}}C(t,r)}{A^3 R^4} \Big[&(A^2 H_5^{(g)} -H_5^{(f)}) ^2-l(l+1) (A^2H_5^{(g)}-H_5^{(f)}) (A^2 H_6^{(g)} - H_5^{(f)}) \notag\\
  &+ \frac{l(l+1)}{2}(A^2 H_6^{(g)} - H_6^{(f)}) ^2\Big],
\end{align}
where $\hat{M}_{\rm pl}^2=M_{\rm pl}^2 \pi /(1+2l)$ . Then, the
following six primary constraints are imposed,
\begin{eqnarray}
 {\cal C}^{(0)}_{I,(g/f)} &:=& P^{(g/f)}_I \approx 0
\end{eqnarray}
for $I=0,1,2$. The total Hamiltonian is
\begin{eqnarray}
  H_T^{\rm even} = H^{\rm even} + \int dr\left[ v^I_{(g)}(t,r)P_I ^{(g)}+ v^I_{(f)}(t,r)P^{(f)}_I  \right].
\end{eqnarray}
Time evolution of primary constraints is
\begin{eqnarray}
  \dot{{\cal C}}^{(0)}_{I,(g/f)} &=& \partial_t {\cal C}^{(0)}_{I,(g/f)}+\{{\cal C}^{(0)}_{I,(g/f)} , H_T \} \approx\{P^{(g/f)}_I , H_{GR,(g/f)}^{\rm even} \}=  {\cal C}^{(1)}_{I,(g/f)},
\end{eqnarray}
which impose six secondary constraints,
\begin{eqnarray}
  {\cal C}^{(1)}_{I,(g/f)} \approx 0.
\end{eqnarray}
Time evolution of these constraints is given by
\begin{eqnarray}
  \dot{{\cal C}}^{(1)}_{0,(g)} &=& -2   \hat{M}_{\rm pl}^2 \frac{\sqrt{-\bar{g}}C(t,r)}{A R^3} \bar{g}^{0B}\partial_B R \Bigl[(A^2 H^{(g)}_5-H^{(f)}_5)-\frac{l(l+1)}{2}(A^2 H_6^{(g)}-H_6^{(f)})\Bigr],\\
  \dot{{\cal C}}^{(1)}_{1,(g)} &=&  -4 \hat{M}_{\rm pl}^2 \frac{\sqrt{-\bar{g}}C(t,r)}{A R^3} \bar{g}^{1B}\partial_B R\Bigl[(A^2 H^{(g)}_5-H^{(f)}_5)-\frac{l(l+1)}{2}(A^2 H_6^{(g)}-H_6^{(f)})\Bigr], \\
  \dot{{\cal C}}^{(1)}_{2,(g)} &=& 2 \hat{M}_{\rm pl}^2 \frac{\sqrt{-\bar{g}}C(t,r)}{A R^4}l(l+1)\Bigl[(A^2 H_5^{(g)}-H_5^{(f)})-(A^2 H_6^{(g)}-H_6^{(f)})\Bigr],
\end{eqnarray}
with $B= 0,1$, and similar terms appear in the constraints for $f_{\mu\nu}$. Consequently, we
obtain two additional constraints:
\begin{eqnarray}
  {\cal C}^{(2)}_{1} &:=& A^2 H_5^{(g)} - H_5^{(f)},\\
  {\cal C}^{(2)}_{2} &:=& A^2 H_6^{(g)} - H_6^{(f)}.
\end{eqnarray}
The time evolutions of these constraints are given by 
\begin{eqnarray}
  \dot{{\cal C}}^{(2)}_{1} &=& A^2\left( R g^{0B}\partial_B R H^{(g)}_0 +2 R g^{1B}\partial_B R H^{(g)}_1\right)
   - \left( A^2 R f^{0B}\partial_B R H^{(f)}_0 +2 A^2 R f^{1B}\partial_B R H^{(f)}_1 \right)\notag\\
  && + \text{linear terms of $H^{(g/f)}_4,H^{(g/f)}_5,H^{(g/f)}_6,P^{(g/f)}_3,P^{(g/f)}_5,P^{(g/f)}_6$},\\
  \dot{{\cal C}}^{(2)}_{2} &=& 2 A^2 H_2^{(g)}  - 2 H_2^{(f)} + \text{linear terms of $H^{(g/f)}_4,H^{(g/f)}_5,H^{(g/f)}_6,P^{(g/f)}_5,P^{(g/f)}_6$}, 
\end{eqnarray}
which impose further two constraints,
\begin{eqnarray}
  {\cal C}^{(3)}_{1} &:=& \dot{{\cal C}}_1^{(2)} \approx 0,\\
  {\cal C}^{(3)}_{2} &:=&  \dot{{\cal C}}_2^{(2)} \approx 0.
\end{eqnarray}
Since the above constraints include $H^{(g/f)}_0,H^{(g/f)}_1,H^{(g/f)}_2$,
time development of these constraints only determines two of the
multipliers $v_{(g/f)}^I$ and hence no more constraint appears. One can
see that four of the multipliers $v_{(g/f)}^I$ remain undetermined, which
implies that this system has corresponding gauge symmetry. Concrete
calculation shows that this system has eight first class constraints (and eight
second class constraints) through eight non-zero eigenvalues of 16 $\times$
16 matrix $\{{\cal C}_I , {\cal C}_J\}$, where ${\cal C}_I$ represent
all of the sixteen constraints.
These eight constraints are composed of six gauge symmetry of full theory and two additional symmetry described by $\Xi$ in eqs.~(\ref{gauge0}) and~(\ref{gauge1}).
To summarize, the number of graviton degrees
of freedom for even modes can be estimated as
\begin{eqnarray}
  \frac{1}{2} \left(\overbrace{28}^\text{variables} - \overbrace{16}^\text{constraints} - \overbrace{8}^\text{gauge dofs}\right) = 2,
\end{eqnarray}
which again coincides with that of two massless gravitons for $l \ge 2$
modes.

The structure of Hamiltonian analysis is similar for $l=0,1$ mode. For
$l=0$ mode, initially we have eight variables (sixteen phase space variables),
$H^{(g/f)}_0,H^{(g/f)}_1,H^{(g/f)}_3,H^{(g/f)}_5$. Similar analysis shows that
there are ten constraints, ${\cal C}^{(0)}_{0,(g/f)}$,${\cal
C}^{(0)}_{1,(g/f)}$, ${\cal C}^{(1)}_{0,(g/f)}$, ${\cal C}^{(1)}_{1,(g/f)}$,
${\cal C}^{(2)}_{1}$, ${\cal C}^{(3)}_{1}$ and six gauge degrees of
freedom, that is, there are six first class constraints and four second class
constraints.
Four gauge degrees of freedom come from the ones of the full theory and two come from the additional ones described by $\Xi$ in eqs.~(\ref{gauge0}),(\ref{gauge1}). 
Then, the number of dynamical degrees of freedom is
\begin{eqnarray}
 \frac{1}{2} \left(\overbrace{16}^\text{variables} - \overbrace{10}^\text{constraints} - \overbrace{6}^\text{gauge dofs}\right) = 0 .
\end{eqnarray}
For
$l=1$ mode, we have twelve variables (twenty-four phase space variables),
$H^{(g/f)}_0$, $H^{(g/f)}_1$, $H^{(g/f)}_2$, $H^{(g/f)}_3$, $H^{(g/f)}_4$
,$H^{(g/f)}_5$, fourteen constraints, ${\cal C}^{(0)}_{0,(g/f)}$, ${\cal
C}^{(0)}_{1,(g/f)}$, ${\cal C}^{(0)}_{2,(g/f)}$, ${\cal C}^{(1)}_{0,(g/f)}$,
${\cal C}^{(1)}_{1,(g/f)}$, ${\cal C}^{(1)}_{2,(g/f)}$, ${\cal
C}^{(2)}_{1}$, ${\cal C}^{(3)}_{1}$ and ten gauge degrees of freedom,
that is, there are ten first class constraints and four second class
constraints. Then, the number of dynamical degrees of freedom is
\begin{eqnarray}
   \frac{1}{2} \left(\overbrace{24}^\text{variables} - \overbrace{14}^\text{constraints} - \overbrace{10}^\text{gauge dofs}\right) = 0 .
\end{eqnarray}
It should be noted that 
six gauge symmetries correspond to the one of full theory, two gauge symmetries correspond to $\Xi$ in eqs.~(\ref{gauge0}),(\ref{gauge1}), and the other two gauge symmetries correspond to $Q(t,r)$ in eq.~(\ref{gauge2}). 
\bibliography{ref}
\end{document}